# Parity-Time Symmetry in Coherently Coupled Vertical Cavity Laser Arrays

Zihe Gao, Stewart T. M. Fryslie, Bradley J. Thompson, P. Scott Carney, and Kent D. Choquette

Department of Electrical and Computer Engineering, University of Illinois at Urbana-Champaign, Urbana, IL 61820

**Abstract**

We report parity-time (PT) symmetry breaking in electrically injected, coherently coupled, vertical cavity surface emitting laser arrays. We predict beam steering, mode evolution and mode hopping as a consequence of the non-Hermiticity of the array analyzed by temporal coupled mode theory with both asymmetric gain distribution and local frequency detuning. We present experimental confirmation of the predicted mode evolution, mode hopping and PT symmetry breaking with quantitative agreement with the theory.

**Introduction**

The optical realization of non-Hermitian parity-time (PT) symmetric system has drawn great interest in rent years [1–16]. An optical system with a symmetric index profile and an antisymmetric gain/loss profile formally exhibits PT symmetry, by an analogy to the PT symmetry in quantum mechanics [1,2]. One realization of this index (and gain/loss) profile is two coupled waveguides or resonators, where one of the waveguides or resonators exhibits optical gain and the other one is lossy [6,13–15]. PT symmetry breaking, exceptional points [6,13–15,17,18], unidirectional light propagation [14], side-mode suppression [13] have been previously experimentally demonstrated. The optical gain in the majority of the systems above is provided by optical pumping. Below we discuss electrically injected diodes which are appealing and necessary for practical applications in integrated photonics. Photonic integrated circuits are currently under development because of their significant performance advantages in many existing and new deployment areas, such as data centers, smart phones, autonomous automobiles, etc. Compared to optically pumped PT-symmetric coupled resonators, an electrically pumped system presents new and interesting physical effects. For example, selective pumping perturbs the index profile through carrier and thermal effect in addition to the gain/loss profile thus generating a frequency detuning. We propose temporal coupled mode theory as a means to analyze and make quantitative predictions of the optically coupled resonator behavior. We apply this theory specifically to an electrically injected, coherently coupled, 1×2 vertical cavity surface emitting laser (VCSEL) array.

While the study of PT-symmetric Hamiltonians traces to the seminal paper in 1998 [19], coherently coupled semiconductor laser arrays have been studied for more than three decades [20–28] for applications including high brightness beam generation [29–32], beam steering [21,27,33,34], and high bandwidth modulation [35,36]. Coupled mode theory has been used to describe the supermodes of the array [23,37,38], phase velocity matching [39], and dynamics [27,40]. However, in previous couple-mode analyses, the possibility of gain difference between the individual resonators was either not included or was implicit in a set of carrier density equations coupled to the optical field equations. The analysis here takes gain in the individual resonators as an explicit parameter (imaginary part of the complex frequency) instead of introducing coupled carrier density equations. This preserves the linearity of the coupled mode theory while still offering insight into salient physical effects.

The effects of gain/loss contrast, frequency detuning, and their interplay on the supermodes of coherent 1×2 VCSEL arrays are discussed. It is shown below that beam steering in coherently coupled lasers is a direct result from the optical non-Hermiticity of the system caused by gain-loss contrast, and that the π/2 limit of the maximum relative phase tuning [21,41,42] is reached at or beyond the exceptional point in the PT-symmetry broken regime. This beam steering mechanism of directly coupled resonators is essentially different from the case where the resonators are locked to a common master [43,44], or in the optical nano-antenna arrays [45]. Experimentally we demonstrate evolution of the lasing supermode, including beam steering, near-field intensity distribution, hopping of the lasing mode between in-phase and out-of-phase modes, and parity-time symmetry breaking. We also show our experimental measurements including operating wavelength, relative intensities, and relative phase agree quantitatively with the temporal coupled mode theory.

**Temporal Coupled Mode Theory with Gain/Loss Contrast Included**

Given two resonant modes that are weakly coupled, the amplitudes $E_a$ and $E_b$ of the two modes can be described by

$$\begin{aligned}\frac{dE_a}{dt} &= -i\omega_a E_a + \gamma_a E_a + i\kappa_{ab} E_b \\ \frac{dE_b}{dt} &= -i\omega_b E_b + \gamma_b E_b + i\kappa_{ba} E_a\end{aligned} \quad (1)$$

where $\omega_{a,b}$ are the natural resonant frequencies of the two modes, $\kappa_{ab,ba}$ are the coupling coefficients, and $\gamma_{a,b}$ are the temporal gain/loss coefficients (positive value represents gain and negative represents loss). Weak coupling means that $\kappa_{ab,ba} \ll \omega_{a,b}$. Here we assume that all the coefficients, $\omega_{a,b}$, $\gamma_{a,b}$, and $\kappa_{ab,ba}$ are constants, which forces the system to be linear and ignores all the nonlinear effects such as gain saturation or dynamic coupling between field amplitudes and carrier densities. This approximation can be justified by either considering only small field amplitudes at or below threshold where gain is not saturated, or by assigning the gain coefficient to be the saturated value given knowledge of the field amplitudes. Equation (1) can be written in compact form as

$$i\frac{d\mathbf{E}}{dt} = \mathbf{\Omega}\mathbf{E},\qquad(2)$$

where $\mathbf{E} = \begin{pmatrix} E_a \\ E_b \end{pmatrix}$ and $\mathbf{\Omega} = \begin{pmatrix} \omega_a + i\gamma_a & -\kappa_{ab} \\ -\kappa_{ba} & \omega_b + i\gamma_b \end{pmatrix}$. Equation (2) has the same form as the Schrödinger equation in quantum mechanics, where $\mathbf{\Omega}$ is an optical analogy to the Hamiltonian of a quantum particle inhabiting a coordinate axis that consists of just two points [46]. The action of the operator $\mathbf{PT}$ as an analogy to the quantum particle can be defined as

$$(\mathbf{PT\Omega})_{i,j} = \mathbf{\Omega}^*_{3-i,3-j},\qquad(3)$$

where $i, j$ are the matrix indices taking the value of $1, 2$, and hence a general PT-symmetric $\mathbf{\Omega}$ can be expressed as $\mathbf{\Omega} = \begin{pmatrix} x & y \\ y^* & x^* \end{pmatrix}$ [46]. In a power-conserving, lossless, gainless, coupled-resonator system, $\kappa_{ab} = \kappa_{ba}^*$ $\gamma_a = \gamma_b = 0$, which means $\mathbf{\Omega}$ is Hermitian (and PT-symmetric if $\omega_a = \omega_b$). If balanced gain and loss are judiciously introduced (i.e. $\gamma_a = -\gamma_b$), the system is not Hermitian, but can still be PT-symmetric if $\omega_a = \omega_b$. Note that the condition of balanced gain and loss, $\gamma_a = -\gamma_b$, can be relaxed to any gain or loss contrast, as long as $\gamma_a \neq \gamma_b$, if we change variables to extract a common gain/loss factor $\exp[(\gamma_1 + \gamma_2)t/2]$ from the temporal dependence of the field amplitudes [5,47].

Demanding that the field amplitudes are time harmonic $\mathbf{E} = \begin{pmatrix} A_a \\ A_b \end{pmatrix} e^{-i\omega t} = \mathbf{A}e^{-i\omega t}$, Equation (2) becomes an eigenvalue problem $\mathbf{\Omega A} = W\mathbf{A}$. Eigenvalue $W$ represents the complex frequency of the coupled modes, with real part representing the angular frequency and imaginary part representing the gain/loss coefficient. The solution of the eigenvalue equation is

$$\omega = \frac{\omega_a + \omega_b}{2} + \frac{i(\gamma_a + \gamma_b)}{2} \pm \sqrt{\kappa_{ab}\kappa_{ba} + (\frac{\omega_a - \omega_b}{2})^2 - (\frac{\gamma_a - \gamma_b}{2})^2 + \frac{i(\omega_a - \omega_b)(\gamma_a - \gamma_b)}{2}} \quad . \tag{4}$$

The eigenvector $\mathbf{A} = \begin{pmatrix} A_a \\ A_b \end{pmatrix}$ represents the composition of the supermodes, which is a superposition of the two individual modes. $\mathbf{A}$ can be calculated from the eigen-frequencies by

$$\frac{A_b}{A_a} = \frac{E_b}{E_a} = \frac{-i\kappa_{ba}}{i(\omega - \omega_b) + \gamma_b} \quad . \tag{5}$$

The magnitude of the ratio, $|A_b / A_a|$, controls the near field intensity profile of the mode, while the phase of the ratio, $Ang(A_b / A_a)$, controls relative phase tuning, which leads to beam steering in the far field [42]. Equations (4) and (5) take simpler forms in specific cases, as discussed below. For simplicity we set the general complex coupling coefficient to be symmetric, real and positive, i.e. $\kappa_{ab} = \kappa_{ba} = \kappa > 0$, assuming negligible deviation in the coupling coefficients from the case of two identical passive resonators.

When the two resonators have different native resonant frequencies but experience the same amount of gain or loss, i.e. $\omega_a \neq \omega_b, \gamma_a = \gamma_b = \gamma$, we have

$$\begin{aligned} \omega &= \frac{\omega_a + \omega_b}{2} + i\gamma \pm \sqrt{\kappa^2 + (\frac{\omega_a - \omega_b}{2})^2} \\ \frac{E_b}{E_a} &= \frac{\omega_a - \omega_b}{2\kappa} \mp \sqrt{1 + (\frac{\omega_a - \omega_b}{2\kappa})^2} \end{aligned} \quad . \tag{6}$$

The frequency splitting between the two coupled modes is at its minimum, $2\kappa$, when the frequency detuning is zero. $E_b/E_a$ is always a real number, which means the relative phase between the field amplitudes in two resonators is either 0 (in-phase mode) or $\pi$ (out-of-phase mode). In this situation, the wavelengths, gain, field amplitude ratio (both magnitude and phase) are illustrated in Fig. 1. The frequency detuning changes the coupled-mode intensity distribution, such that the out-of-phase mode has more intensity in the resonator with higher natural resonant frequency, while the in-phase mode has more intensity in the cavity with lower natural resonant frequency. The degree of intensity distribution asymmetry increases with frequency detuning as shown in Fig. 1(c). The fact that the eigenvector $\mathbf{A}$ is purely real for a Hermitian $\mathbf{\Omega}$ means no phase tuning or beam steering is induced if gain contrast is absent, as shown in Fig. 1(d).

When the two resonators have identical native resonant frequency, but experience different amount of gain or loss i.e. $\omega_a = \omega_b = \omega_0, \gamma_a \neq \gamma_b$, $\mathbf{\Omega}$ is not Hermitian but PT symmetric. In this case the complex frequencies of the eigenmodes are

$$\omega = \omega_0 + \frac{i(\gamma_a + \gamma_b)}{2} \pm \sqrt{\kappa^2 - (\frac{\gamma_a - \gamma_b}{2})^2} \ . \qquad (7)$$

If the square root is real in Equation (7), the coupled system is in the unbroken PT symmetry regime and $A_b/A_a = \mp exp(\mp i\theta)$ $\sin\theta = (\gamma_a - \gamma_b)/2\kappa$. In the unbroken PT symmetry regime, the coupled modes are balanced superpositions of the individual modes ($|A_b/A_a| = 1$), and relative phase tuning is achieved by varying the gain contrast $\gamma_a - \gamma_b$. On the other hand, if the square root is imaginary in Equation (7), the coupled system is in the broken PT symmetry regime and $A_b/A_a = i\exp(\mp\theta)$ $\cosh\theta = \frac{\gamma_a - \gamma_b}{2\kappa}, \theta > 0$ [6,17]. In the broken PT symmetry regime the coupled modes are unbalanced superpositions of the individual modes, and the relative phase between the two cavities is fixed at the value of ±π/2. The evolution of the coupled modes with varying gain contrast are illustrated in Fig. 2.

The fact that the phase tuning cannot exceed the limit of ±π/2 has been reported in coupled laser arrays experiments; see for example [21,41,42]. Upon hitting the π/2 phase tuning limit, further increasing of the gain contrast results in driving the array into PT symmetry broken regime, where the relative phase is pinned at π/2 and the intensity distribution of the coupled modes become asymmetric. It has been observed in previous experiments that the mutual coherence between the cavities decreases when the π/2 phase tuning limit is reached [41,42]. This loss of mutual coherence can be expected as the coupled modes become asymmetric and spatially concentrate in single cavities, resulting in the simultaneous lasing of both coupled modes. Another feature worth noting in Fig. 2 is the appearance of exceptional points or branching points at $\Delta g / \kappa = \pm 2$, points at which the two eigenmodes collapse. Because of the collapse of eigenmodes, in Fig. 2(b) and Fig. 2(c) the modes are not labeled as in-phase or out-of-phase to avoid confusion, because unlike the case in the Fig. 3, the coupled modes cannot be traced back across the exceptional point to be identified as in-phase or out-of-phase modes.

When both gain contrast and frequency detuning exist, the coupled modes are controlled by the interplay of frequency detuning and gain contrast. In Fig. 3, frequency detuning is assumed to be proportional to the amount of gain contrast, both linearly dependent on the injected current difference [48,49]. The degeneracies we see in the ideal PT-symmetric case (Fig. 2) do not exist when frequency detuning is present. Also note that the gain of the (skewed) in-phase mode is higher than the (skewed) out-of-phase mode when the current injection difference is nonzero. This is because the change of intensity distribution of the in-phase mode, as a result of simultaneous frequency detuning and gain contrast, enhances its spatial overlap with the spatially non-uniform gain, while the intensity distribution change of the out-of-phase mode does the opposite. Whether it is the in-phase mode or out-of-phase mode that gets higher gain depends on the sign of $(\omega_a - \omega_b)(\gamma_a - \gamma_b)$. For the out-of-phase mode to have higher gain requires the local resonant frequency to increase with increasing local gain, for example if carrier induced index suppression dominates the thermal effect. It has been known that evanescently coupled VCSEL arrays tend to work in out-of-phase mode due to less spatial overlap with the lossy inter-element area, although for most applications in-phase mode is preferred. The gain discrimination preference for the in-phase mode suggests that with sufficiently large current injection difference, the mode may hop from out-of-phase mode to in-phase mode, offering a novel modal control method and reconfigurability. This mode hopping behavior is observed experimentally and discussed in the experimental section below. Fig. 3(d) also shows that the phase tuning limit is smaller than π/2. It indicates that to achieve the theoretical limit of π/2 phase tuning, one need to minimize the frequency detuning accompanying non-uniform pumping.

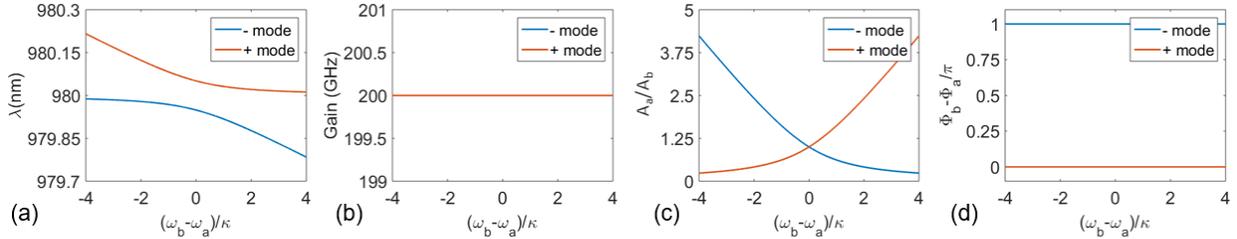

Fig. 1. Effect of frequency detuning without gain contrast on (a) wavelengths of the coupled modes (b) gain of the coupled modes (c) ratio of the field magnitudes in two cavities (d) relative phase between the field in two cavities.

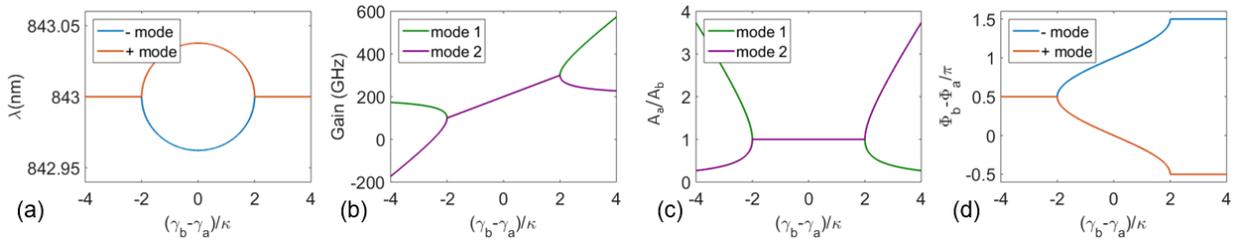

Fig. 2. Effect of gain contrast without frequency detuning on (a) wavelengths of the coupled modes (b) gain of the coupled modes (c) ratio of the field magnitudes in two cavities (d) relative phase between the field in two cavities.

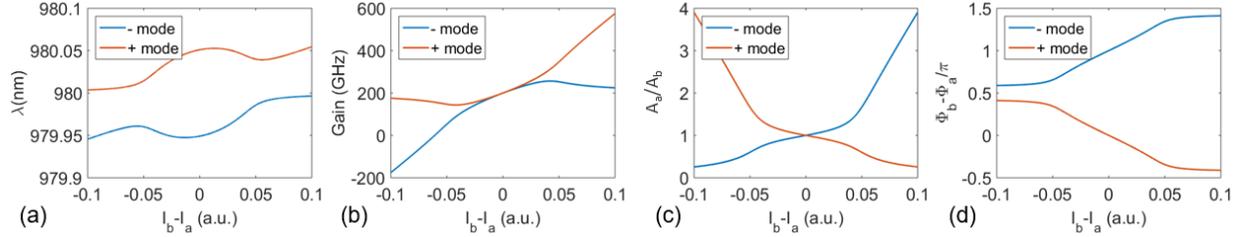

Fig. 3. Effect of co-existing gain contrast and frequency detuning on (a) wavelengths of the coupled modes (b) gain of the coupled modes (c) ratio of the field magnitudes in two cavities (d) relative phase between the field in two cavities. It is assumed that the local changes of gain and frequency are linearly dependent on the current difference, with $\Delta g = -4\Delta\omega$, and the maximum gain contrast at the edge of the graphs are $\Delta g_{max} = 2\kappa$.

## Characterization of Coherently Coupled Photonic Crystal VCSEL Arrays

The above analysis can be applied to $1 \times 2$ coherently coupled photonic crystal VCSEL arrays, shown in Fig. 4(a) [28,32]. The electrical apertures of the VCSELs are defined by proton ion implantation, and a focused ion beam (FIB) etch between the two cavities improves the electrical isolation enabling individual control of injected currents [50]. Lateral optical confinement is provided by photonic crystal patterns etched into top facet leading to single mode operation for each element VCSEL. Details of the device structure and fabrication process have been described in [28]. In coupled VCSEL arrays, the eigenvector **A** of the lasing coupled mode can be characterized from near field and far field measurements, as shown in Fig. 4(b) and (c) [42]. The relative phase between the two cavities can be extracted by propagating the near field with the assumed phase, checking the propagated far field with the measured far field, and iterating with a better assumed phase if far fields do not agree [42]. Fig. 4 (d) shows the agreement of the propagated and measured far fields.

Fig. 5 is a summary of the measured coupled mode wavelength, near field amplitude ratio, and extracted relative phase compared with the coupled mode theory. As shown in the bottom of Fig. 4(b), (c), and (d), the lasing mode starts as a slightly skewed out-of-phase mode at equal injected currents $I_a = I_b = 3.99$mA. Then as $I_b$ is decreased, from $I_b = I_a$ to $I_b = 3.87$mA, we see the minimum of the far field moves away from the center, as a result of beam steering. At $I_b = 3.86$mA ($\Delta I = -0.13$mA) the lasing mode hops to the (skewed) in-phase mode, indicated by the abrupt change in the relative phase (Fig. 5(d)), wavelength (Fig. 5(a)), and the near field intensity distribution (Fig. 5(c)), marked by the red arrows in Fig. 5. $I_b$ was then decreased to 3.83mA and increased back to 3.99mA, the reverse mode hopping happened at $\Delta I = -0.08$mA, marked by the blue arrows in Fig. 5, showing hysteresis and bistability in the region of $-0.13$mA $< \Delta I < -0.08$mA. We cannot measure gain of the coupled modes directly but the mode hopping is an indication of the change in gain of the respective modes. The measured intensity ratio data appears to be offset from the theory in Fig. 5(c). This is because the two elements were not identical and the output of laser *a* was stronger than laser *b* even with the same injected currents, which is believed to be a result of fabrication imperfections (for example the slightly off-center FIB etch).

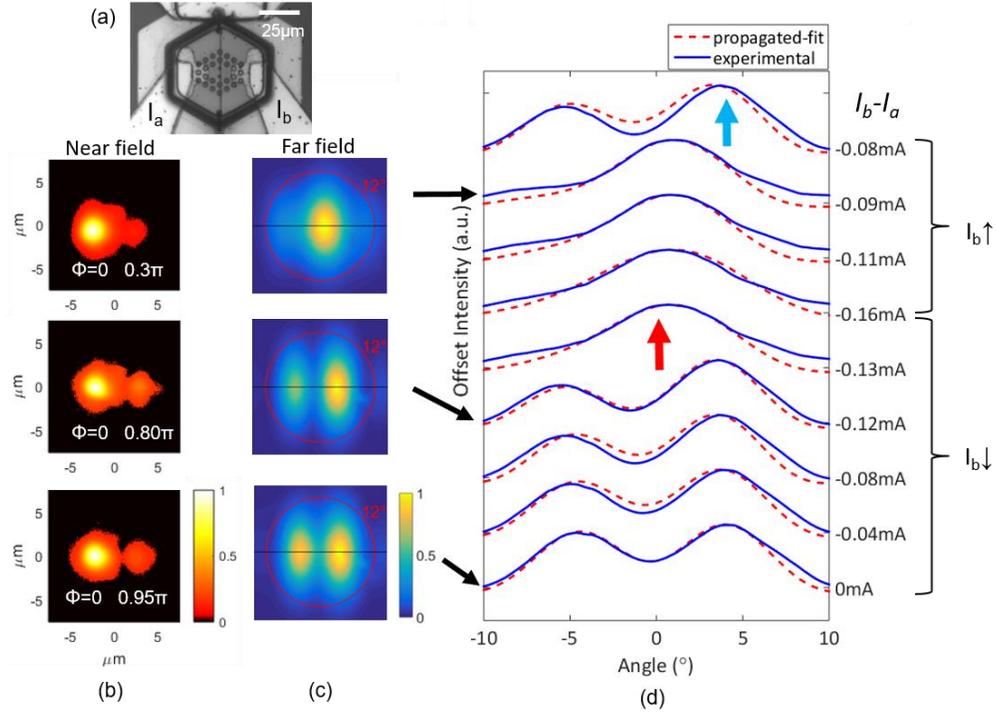

Fig. 4. (a) Microscopic photo of the device under characterization; (b) Near field intensity profiles with extracted relative phase [42], and (c) far field intensity profiles measured at ΔI = 0mA, -0.12mA, -0.09 mA respectively; (d) Evolution of the far field profile with varying $I_b$, with both measured far fields and the ones from propagating the near fields.

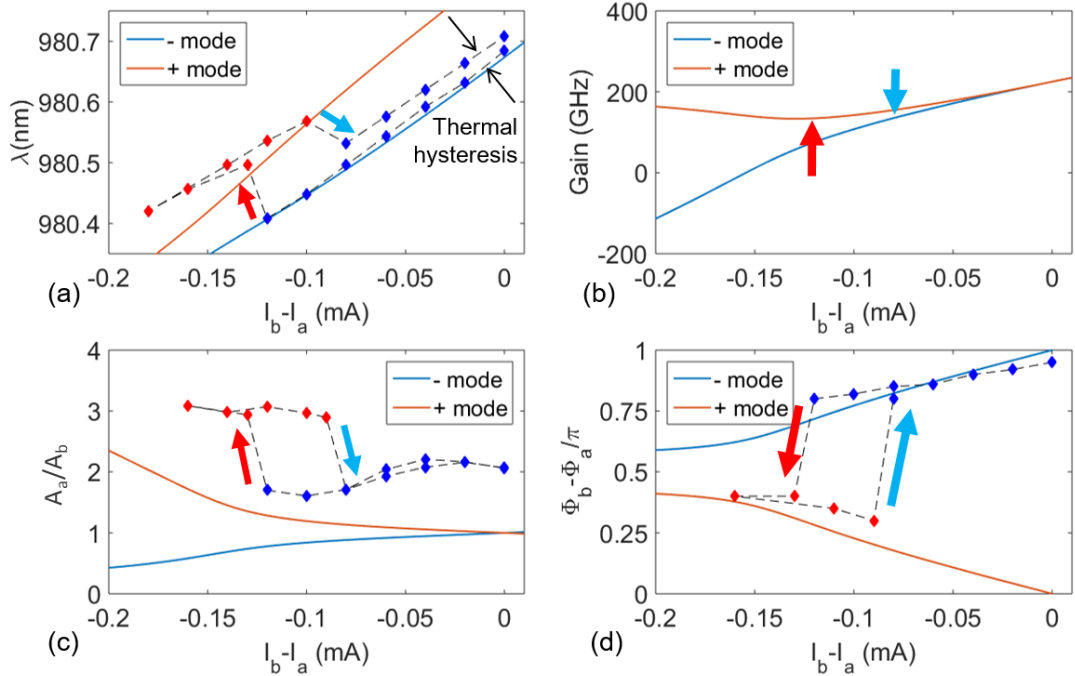

Fig. 5. Comparison of measured data (diamond points) with coupled mode theory (solid lines) on (a) wavelength, (b) gain, (c) field magnitude ratio, and (d) relative phase. Parameters used in coupled mode theory are: $\kappa = 1.5 \times 10^{11}\, Hz$, maximum gain contrast and frequency detuning at $I_b - I_a$ = -0.2mA are $\Delta g_{max} = 1 \times 10^{12}\, Hz$ and $\Delta \omega_{max} = 2 \times 10^{11}\, Hz$ respectively. The two cavities are assumed to have the same resonant wavelength at $I_b - I_a = 0$ and both redshift when $I_b$ is decreased, with $\omega_b$ shifting more than $\omega_a$. Arrows indicate mode hopping.

As discussed previously, PT symmetry breaking can be identified by the pinning of the phase to π/2. For the device shown in Fig. 4 and Fig. 5, the maximum relative phase appears to be limited to 0.4π as a result of the frequency detuning expected to accompany selective electrical pumping. If the component VCSELS initially exhibited a native frequency splitting that was eliminated by the effects of selective electrical pumping, we would expect to see a closer approach to the theoretical maximum phase difference. For another VCSEL array that we have characterized, the π/2 relative phase was obtained, with near field and far field intensity profiles shown in Fig. 6(a) and (b), showing good agreement with the simulated broken PT-symmetry eigenstate in Fig. 6(c) and (d). For this particular coupled array, as a result of fabrication imperfection the two VCSEL elements are spectrally aligned when the gain contrast was large enough for PT symmetry breaking.

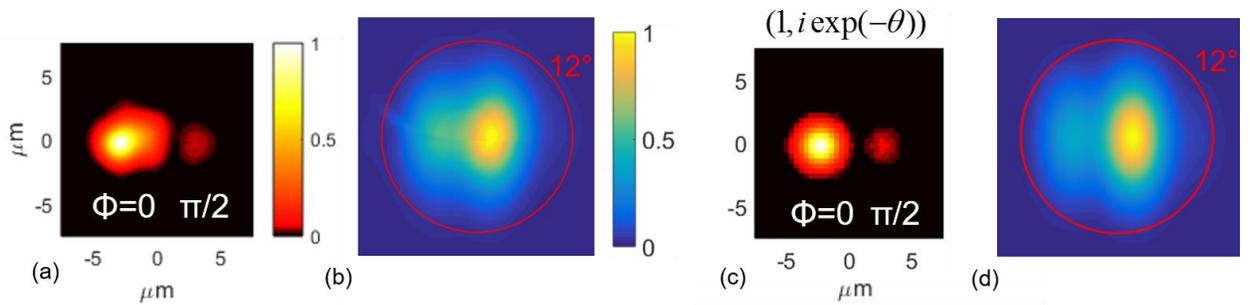

Fig. 6. Measured (a) near field and (b) far field in PT symmetry broken regime compared with the simulated (c) near field and (d) far field of the $(1, i\exp(-\theta))$ eigenmode.

**Conclusions**

In summary, electrically injected optically coupled VCSEL arrays are well-modelled using time-domain coupled mode theory, allowing us to account for gain contrast and frequency splitting, and predicting a range of physical behaviors associated with PT symmetry and symmetry breaking. We have observed PT-symmetry-related mode evolution and PT-symmetry breaking in room temperature coupled VCSEL diode arrays with quantitative agreement with the theory. A controllable mode hopping between in-phase and out-of-phase mode is identified here for the first time. This work demonstrates the potential of designing the gain/loss profile for non-Hermitian engineering of coupled optical systems and may lead to new engineering applications in the future.

**Acknowledgements**

The authors thank M. P. Tan, D. Siriani, and M. T. Johnson. This material is based upon work partially supported by the National Science Foundation under Award No. ECCS 15-09845. ZG and KDC thank M. Khajavikhan and V. Kovanis for discussions.